\documentclass[%
 reprint,
amsmath,amssymb,
aps,
prl,
longbibliography
]{revtex4-2}

\usepackage[pdftex]{graphicx}

\graphicspath{
{./Figures/},
}

\usepackage{chemformula} 
\usepackage{diagbox}
\usepackage{braket}
\usepackage{dcolumn}
\usepackage{bm}
\usepackage{booktabs}
\usepackage{threeparttable}

\begin{document}

\preprint{APS/123-QED}

\title{Mathematical crystal chemistry: A formal theory for crystal structure prediction by generalized disjunctive programming}

\author{Ryotaro Koshoji}
\affiliation{Institute for Solid State Physics, The University of Tokyo, Kashiwa 277-8581, Japan}
\email{cosaji@issp.u-tokyo.ac.jp}

\date{\today}

\begin{abstract}

\noindent
Inorganic structural chemistry leads naturally to a theory for crystal structure prediction formalized by a generalized disjunctive programming (GDP),
which is formulated using continuous and Boolean variables to involve the algebraic equations, disjunctions, and logic propositions.
Since the feasibilities of continuous variables change drastically depending on Boolean variables and vice versa,
iterative optimization of continuous and Boolean variables efficiently transforms a randomly generated initial structure into an optimal solution.
Boolean variables are introduced to elucidate the ``combinatorial backbone'' of the original continuous optimization problem,
which corresponds to the graphs describing crystal structures with clear chemical meanings.
Since many subgraphs are discovered in numerous graphs generated through structural optimizations,
on-the-fly learning of the feasibilities of them accelerates the discovery of the optimal solutions.
This theory enables designing a wide variety of crystal structures with small computations based on only the atomic radii and feasible coordination numbers of each atom.

\end{abstract}

\maketitle

Crystal structure prediction is the optimization problem of the Gibbs free energy,
which is the function of continuous variables including spatial arrangement of atoms through the many-body wavefunction.
This is a very difficult optimization problem like a long-standing scandal in physics~\cite{Maddox1988} due to countless local optima,
but many techniques using metaheuristics such as random search methods~\cite{PhysRevLett.97.045504, Pickard_2011, D4FD00134F},
particle swarm optimization~\cite{PhysRevB.82.094116, WANG2016406, 10.1063/5.0074677, Luo2024},
or evolutionary algorithms~\cite{10.1039/9781788010122, 10.1063/1.2210932, Liu2021, Kruglov2023, doi:10.1021/acs.jpcc.0c09531, HAJINAZAR2024109306, D4DD00269E, HAJINAZAR2026109910},
have made it now possible to predict a crystal structure given only its composition for systems that are not too complicated.
They are designed to focus on the set of reasonable possibilities for arrangements of atoms,
which is much less than the set of all possible structures.

Essentially, deterministic algorithms to solve optimization problems must visit every local optimum,
but conventional nonlinear optimization techniques for real variables,
which rely on such concepts as derivatives and gradients,
have been mainly concerned with finding local optima.
While there are methods to replace a too difficult original problem with a more easily solvable relaxed problem~\cite{Tomiyasu2018},
no deterministic polynomial time algorithm to solve nondeterministic polynomial (NP) hard problems has been found consistent with $\mathrm{P} \neq \mathrm{NP}$ conjecture.
A common mechanism behind NP-hardness of nonconvex continuous optimization is that the continuous feasible region or objective contains a ``combinatorial backbone'',
such as vertices of a polytope,
or sign patterns,
etc.
The global optimization over the continuous space contains the solution of such an NP-hard discrete problem,
and therefore becomes hard~\cite{doi:10.1287/moor.22.3.754}.
In contrast to the formal theory,
the Branch and Bound algorithm for integer programmings,
which is capable of rapidly eliminating large parts of the optimization domain from consideration,
is devised to find the global optimum more efficiently than random samplings~\cite{Gusev2023}.
Then,
the next question is whether we can introduce integer variables to not only clarify the mathematical structure of crystal structure prediction such as NP-hardness but also suggest efficient algorithms for handling it.

Almost all the eigenvalue problems for many-body wavefunctions are unsolvable.
The density functional theory (DFT),
which replaces them with different auxiliary systems that can be solved more easily,
makes large-scale crystal structure searches possible,
but the exact form of the exchange-correlation energy is unknown.
That is,
nobody knows the explicit mapping spatial orders of atoms to the objective function for crystal structure prediction.
However,
experimental synthesizers of inorganic materials know the empirical rules to elucidate the structures of chemical compounds,
which are organized as the science named inorganic structural chemistry~\cite{doi:10.1021/ja01379a006, StructuralInorganicChemistry, InorganicStructuralChemistry, IntroductionToStructuralChemistry}.
It deals with both the continuous and discrete aspects of crystal structures;
the former is the packing of atomic spheres,
and the latter is the coordination polyhedra and the linking of them.

It was not until 2000s that Kepler conjecture was proved positively despite the simpleness of the problem~\cite{10.2307/20159940}.
The densest packings of spheres of only two or three sizes are still unknown;
however,
many methods have predicted them by computer~\cite{PhysRevE.79.046714, doi:10.1021/jp804953r, doi:10.1021/jp206115p, doi:10.1021/jp1045639, Hudson_2011, PhysRevLett.107.125501, PhysRevE.85.021130, PhysRevE.103.023307, PhysRevE.104.024101, koshoji2021diverse}.
They were applied to design unknown crystal structures such as quaternary hydrides under high pressure~\cite{PhysRevMaterials.6.114802},
but the predictive power from them seems to be limited~\cite{PhysRevE.103.023307, PhysRevE.104.024101, koshoji2021diverse, PhysRevMaterials.6.114802}.

A lot of program packages have been developed to represent the discrete aspect of the structures of chemical compounds by graphs~\cite{doi:10.1021/cg500498k, DELGADOFRIEDRICHS20052533, doi:10.1021/ar800124u, doi:10.1021/cr200205j, 10.1063/1.4901292, doi:10.1021/acs.cgd.3c01492, D5CE01176K}.
Many AI techniques also use the graph neural networks to provide a prediction of not only material properties~\cite{PhysRevLett.120.145301, doi:10.1021/acs.chemmater.9b01294, PhysRevMaterials.4.063801, Jiang2021, Fung2021, Cheng2021, Choudhary2021, Reiser2022} but also crystal structures~\cite{Long2021, doi:10.1126/sciadv.abi7948, Cheng2022, Li2023, Xiao2023, doi:10.1021/acs.jcim.4c01689, Hong_2025, 10.1063/5.0278803}.
However, it is still a challenge to inversely design solid-state materials with graphs,
because the reconstruction of the original crystal structure from the graph is difficult.
Infinite periodic 3D structures with translational,
rotational,
and permutational invariances cannot be represented as undirected finite graphs unlike molecular inverse design~\cite{Xiao2023}.

This letter shows that the next logical step would be to combine the continuous and discrete aspects to develop an efficient solution method for crystal structure prediction.
It is not very difficult to do:
Inorganic structural chemistry leads naturally to a formal theory using continuous and discrete variables,
which not only elucidates the mathematical structure of crystal structure prediction but also provides a high-speed algorithm to design crystal structures.

First, we define the feasible atomic environment of each atom as
\begin{equation}
\left( r^{\left( G \right)}, R^{\left( G \right)}, c^{\left( G \right)}, C^{\left( G \right)} \right) \qquad \text{($\forall G \in \mathcal{G}$)},
\label{eq:atomic-environment-feature}
\end{equation}
where $r^{\left( G \right)}, R^{\left( G \right)} \in \mathbb{R}$ are the minimum and maximum radii, respectively,
$c^{\left( G \right)}, C^{\left( G \right)} \in \mathbb{N}$ are the minimum and maximum coordination numbers, respectively,
and $G \in \mathcal{G}$ correponds to a type of interatomic forces such as ionic bond, covalent bond, and electrostatic repulsion.
If $G$ does not correspond to a chemical bond,
it is quite natural to set $R^{\left( G \right)} = \infty$, $c^{\left( G \right)} = 0$, and $C^{\left( G \right)} = \infty$.
Think about atom $j \bm{T}$ as the atom $j$ in lattice $\bm{T} \in \mathbb{Z}^3$.
To constrain the feasible interatomic distance between atoms $i$ and $j \bm{T}$,
we must select the type of interatomic distance constraints $G$ between them.
At this point, we need to introduce the Boolean variables $Y_{ij \bm{T}} ^{\left( G \right)} \in \left\{ \, \mathrm{True}, \mathrm{False} \, \right\}$,
where $Y_{ij \bm{T}} ^{\left( G \right)}$ is True if and only if $G$ is selected.
We can now write the constraints on interatomic distance:
\begin{equation}
\underset{G \in \mathcal{G}}{\vee} \begin{bmatrix}
Y_{ij \bm{T}}^{\left( G \right)} \\
d_{ij} ^{\left( G \right)} \le \left| \bm{x}_j + A \bm{T} - \bm{x}_i \right| \le D_{ij} ^{\left( G \right)}
\end{bmatrix},
\label{eq:disjunction-term}
\end{equation}
where $d_{ij} ^{\left( G \right)}$ and $D_{ij} ^{\left( G \right)}$ fix the feasible range of interatomic distance,
$A \equiv \left( \bm{a}_1 , \bm{a}_2 , \bm{a}_3 \right)$ is the lattice vectors of the unit cell,
and $\bm{x}_i$ is the position vector of atom $i$.
Each term of disjunction,
which is associated with a Boolean variable and the inequality constraint on interatomic distance,
is linked together by OR operator ($\vee$).
When a disjunctive term is selected ($Y_{ij \bm{T}}^{\left( G \right)}$ is True),
the corresponding inequalities are enforced.
Otherwise, the constraints are ignored.
Since every pair of atoms should select one type of interatomic forces,
we impose the constraints that exactly one of the Boolean variables can be True:
\begin{equation}
\underset{G \in \mathcal{G}}{\veebar} Y_{ij \bm{T}}^{\left( G \right)},
\label{eq:exclusive-or-term}
\end{equation}
where $\veebar$ is the exclusive OR (XOR) operator.
Now recall that each atom has the feasible coordination numbers with respect to $G$.
We arrive at the constraint on the collection of active Boolean variables,
which is represented by cardinality clauses defined by the following predicates:
$\mathrm{Exactly} \left( m, Y_{s}, \; \forall s \in \mathcal{S} \right)$ enforces that exactly $m$ of the Boolean variables $Y_s$ are True,
$\mathrm{AtLeast} \left(m, Y_s \; \forall s \in \mathcal{S} \right)$ enforces that at least $m$ of $Y_s$ are True,
and $\mathrm{AtMost} \left( m, Y_{s}, \; \forall s \in \mathcal{S} \right)$ enforces that at most $m$ are True.

We are now ready to construct the theory for crystal structure prediction.
From inorganic structural chemistry,
we replace the original optimization problem of crystal structure prediction,
which is too difficult to solve,
by the different auxiliary problem that relies on the intersection of disjunctions of algebraic constraints (equality and inequality constraints with continuous variables) controlled by Boolean variables.
Such a model is known as generalized disjunctive programming (GDP)~\cite{MixedIntegerNonlinearProgramming, Grossmann_2021, doi:10.1021/ie2030486, https://doi.org/10.1002/aic.14088, https://doi.org/10.1002/cite.201400037, Perez2024, doi:10.1021/acs.iecr.5c02853}.
GDP can include logic propositions expressed by logic operators such as AND ($\land$),
OR ($\vee$),
XOR ($\veebar$),
negation ($\neg$),
implication ($\implies$),
and equivalence ($\iff$).
We are now ready to write down the GDP to search crystal structures.
Here, it is:
\begin{widetext}
\begin{equation}
\begin{split}
\underset{ A, \; \bm{x}_i, \: Y_{ij \bm{T}}^{\left( G \right)}}{\text{minimize}} \qquad & \left| \det{A} \right| \\
\text{subject to} \qquad & \mathrm{AtLeast} \left( c_i ^{\left( G \right)}, \, Y_{ij \bm{T}} ^{\left( G \right)}, \;\; \forall j \in \mathcal{I}, \forall \bm{T} \in \mathbb{Z}^3 \right) \hspace{3.5em} \text{($\forall i \in \mathcal{I}$, $\forall G \in \mathcal{G}$)} \\
& \mathrm{AtMost} \left( C_i ^{\left( G \right)}, Y_{ij \bm{T}} ^{\left( G \right)}, \;\; \forall j \in \mathcal{I}, \forall \bm{T} \in \mathbb{Z}^3 \right) \hspace{3.45em} \text{($\forall i \in \mathcal{I}$, $\forall G \in \mathcal{G}$)} \\
& \underset{G \in \mathcal{G}}{\veebar} Y_{ij \bm{T}}^{\left( G \right)} \hspace{16.65em} \text{($\forall i,j \in \mathcal{I}$, $\forall \bm{T} \in \mathbb{Z}^3$)} \\
& \underset{G \in \mathcal{G}}{\vee} \begin{bmatrix}
Y_{ij \bm{T}}^{\left( G \right)} \\
d_{ij \bm{T}}^{\left(G \right)} \le \left| \bm{x}_j + A \bm{T} - \bm{x}_i \right| \le D_{ij \bm{T}}^{\left(G \right)}
\end{bmatrix} \hspace{4.1em} \text{($\forall i,j \in \mathcal{I}$, $\forall \bm{T} \in \mathbb{Z}^3$)} \\
& \neg \Lambda _{ij} ^{\left( G \right)} \implies \neg Y_{ij \bm{T}}^{\left( G \right)} \hspace{12.75em} \text{($\forall i,j \in \mathcal{I}$,  $\forall \bm{T} \in \mathbb{Z}^3$, $\forall G \in \mathcal{G}$)} \\
& \Xi _{ij} ^{\left( G \right)} \, \implies \, Y_{ij \bm{T}}^{\left( G \right)} \hspace{13.75em} \text{($\forall i,j \in \mathcal{I}$,  $\forall \bm{T} \in \mathbb{Z}^3$, $\forall G \in \mathcal{G}$)} \\
& d_{ij \bm{T}} ^{\left(G \right)} \equiv r_i ^{\left(G \right)} + r_j ^{\left(G \right)}, \quad\; D_{ij \bm{T}} ^{\left(G \right)} \equiv R_i ^{\left(G \right)} + R_j ^{\left(G \right)} , \quad\; \bm{0} \le A ^{-1} \bm{x}_i \le \bm{1} \\
& A \in \mathbb{R}^3 \times \mathbb{R}^3, \quad\; \bm{x}_i \in \mathbb{R}^3, \quad\; Y_{ij \bm{T}}^{\left( G \right)} \in \left\{ \, \mathrm{True}, \mathrm{False} \, \right\}
\label{eq:gdp-crystal-structure-prediction}
\end{split}
\end{equation}
\end{widetext}
The optimal solutions (prototypes of crystal structures) of the auxiliary GDP must include all the meaningful optimal solutions of the original optimization problem.
Note that since we are mainly interested in questions whether there are feasible solutions satisfying all the constraints,
the auxiliary GDP is essentially a feasibility problem whose objective function is identically zero.
However,
the numerical performance indicates that the volume of unit cell is convenient as the objective function to design crystal structures from random sampling.
Since the GDP model has the concept that it is an extension of inorganic structural chemistry that emphasizes the importance of the closest packings of spheres,
it is reasonable to choose the objective function as it is.
Besides,
since inorganic structural chemistry compiles the tables of atomic radii depending on the types of interatomic force to estimate interatomic distances by the sum of atomic radii~\cite{StructuralInorganicChemistry, InorganicStructuralChemistry, IntroductionToStructuralChemistry, SHANNON:a12967, B801115J},
the GDP model constrains the minimum and maximum interatomic distances by the sum of minimum and maximum atomic radii,
respectively,
to find crystal structures as feasible solutions:
\begin{align}
d_{ij} ^{\left( G \right)} = r_{i}^{\left(G \right)} + r_{j}^{\left(G \right)}, && D_{ij} ^{\left( G \right)} = R_{i}^{\left(G \right)} + R_{j}^{\left(G \right)}
\end{align}
Finally,
the Boolean constants $\Lambda _{ij} ^{\left( G \right)}$ and $\Xi _{ij} ^{\left( G \right)}$ in the logic propositions constrain the possible choices of $G$ between atoms $i$ and $j \bm{T}$.
For example,
if both the two atoms have the same positive formal charge,
an ionic bond is not creatable.
Then,
we can constrain the rule to create ionic bonds as
\begin{equation}
\Lambda _{ij} ^{\left( G \right)} =
\begin{cases}
\mathrm{True} & \text{$q_i q_j < 0$} \\
\mathrm{False} & \text{otherwise}
\end{cases}, \label{eq:ionic_bondable_rule}
\end{equation}
where $q_i$ is the formal charge of atom $i$.

There is no standard definition to quantify chemical bonding properties consistent with chemical intuition,
since chemical bond is certainly not quantum-mechanical observables.
One of the reasonable definitions is given in the software \textsf{LOBSTER}~\cite{Dronskowski+2023, doi:10.1021/acs.jpcc.1c00718, D5SC02936H},
which explains that a covalent bond among atoms is formed by interfering wavefunctions consisting of their overlapping orbitals,
while an ionic bond is formed by electrostatic interactions between charged ions~\cite{Dronskowski+2023, doi:10.1021/acs.jpcc.1c00718, D5SC02936H}.
While covalency and ionicity are often referred to as opposing bonding concepts,
all the chemical bonds in chemical compounds mix both the two concepts.
Therefore,
it seems unreasonable to distinguish the types of interatomic forces,
but the constraint given in Eq.~\eqref{eq:atomic-environment-feature} can include the mixtures of the concepts.
In fact,
since Alkali metals form ionic bonds with very minor covalency,
both the ionic radius and coordination numbers are very flexible.
On the other hand,
transition elements form ionic bonds with strong covalency,
the ionic radius is nearly fixed and the stable coordination numbers are limited.

In general, GDPs can be reformulated as mixed-integer nonlinear programmings (MINLP)s~\cite{https://doi.org/10.1002/aic.690490714, MixedIntegerNonlinearProgramming, Grossmann_2021, doi:10.1021/ie2030486, https://doi.org/10.1002/aic.14088, https://doi.org/10.1002/cite.201400037}.
To do this,
binary variables $y_{ij \bm{T}}^{\left( G \right)} \in \left\{ 0,1 \right\}$ are introduced to have one-to-one correspondences with the Boolean variables:
$Y_{ij \bm{T}}^{\left( G \right)} = \mathrm{True}$ is equivalent to $y_{ij \bm{T}}^{\left( G \right)} = 1$,
while $Y_{ij \bm{T}}^{\left( G \right)} = \mathrm{False}$ is equivalent to $y_{ij \bm{T}}^{\left( G \right)} = 0$.
It is straightforward to convert logic propositions into algebraic forms using binary variables.
The reformulation of Eq.~\eqref{eq:gdp-crystal-structure-prediction} as a MINLP is given in Supplemental Material~\cite{SupplementalMaterial}.
Compared to general MINLPs,
the remarkable feature of the auxiliary GDP given in Eq.~\eqref{eq:gdp-crystal-structure-prediction} is that all the integer variables,
which are equivalent to Boolean variables,
are only used for indicating whether certain constraints on real variables are required to be satisfied or not.

There are two major solution methods for GDP,
namely,
GDP Branch and Bound method and Logic Based Outer-Approximation~\cite{https://doi.org/10.1002/cite.201400037}.
The former directly branches on the disjunctions to generate nodes selecting enforced inequalities.
This procedure is repeated until the tree search is exhausted;
if a node is infeasible, then it is pruned.
The advantage of this method is that it requires the evaluation of fewer nodes than the conventional Branch and Bound methods.
On the other hand,
the latter iteratively solves the subproblem and the master problem.
The subproblem fixes all the Boolean variables to optimize the real variables.
If the optimal solution of the subproblem is discovered,
the collection of the optimized real variables and the fixed Boolean variables is the optimal solution of the original GDP.
If not,
we solve the master problem,
which is aimed at getting a different activation of Boolean variables from the past,
where all the differentiable functions of real variables are linearized by their gradients to optimize both the real and Boolean variables.

Now we will begin to develop an efficient solution method for the auxiliary GDP given in Eq.~\eqref{eq:gdp-crystal-structure-prediction} based on the two conventional solution methods.
As the Logic Based Outer-Approximation,
we create two subproblems,
which are iteratively solved.
One is the optimization problem of the real variables $A$ and $\bm{x}_i$ for the fixed Boolean variables $Y_{ij \bm{T}}^{\left( G \right)}$ derived from Eq.~\eqref{eq:gdp-crystal-structure-prediction}:
\begin{equation}
\begin{split}
\underset{ A, \; \bm{x}_i}{\text{minimize}} \qquad & \left| \det{A} \right| \\
\text{subject to} \qquad & \underset{G \in \mathcal{G}}{\vee} \begin{bmatrix}
Y_{ij \bm{T}}^{\left( G \right)} \\
d_{ij \bm{T}}^{\left(G \right)} \le x_{ij \bm{T}} \le D_{ij \bm{T}}^{\left(G \right)}
\end{bmatrix} \\
& d_{ij \bm{T}} ^{\left(G \right)} \equiv r_i ^{\left(G \right)} + r_j ^{\left(G \right)} \\
& D_{ij \bm{T}} ^{\left(G \right)} \equiv R_i ^{\left(G \right)} + R_j ^{\left(G \right)} \\
& x_{ij \bm{T}} \equiv \left| \bm{x}_j + A \bm{T} - \bm{x}_i \right| \\
& \bm{0} \le A ^{-1} \bm{x}_i \le \bm{1} \\
& A \in \mathbb{R}^3 \times \mathbb{R}^3, \quad\; \bm{x}_i \in \mathbb{R}^3
\label{eq:algebraic-gdp-subproblem}
\end{split}
\end{equation}
The other is the optimization problem of Boolean variables $Y_{ij \bm{T}}^{\left( G \right)}$ for the fixed real variables $A$ and $\bm{x}_i$,
which is indentical to the integer programming:
\begin{widetext}
\begin{equation}
\begin{split}
\underset{b_i ^{\left( G \right)}, \: Y_{ij \bm{T}}^{\left( G \right)}}{\text{maximize}} \qquad & \sum_{i \in \mathcal{I}} \sum_{G \in \mathcal{B}} b_i ^{\left( G \right)} \\
\text{subject to} \qquad & \mathrm{ML} \left( \bm{Y} \right) = \mathrm{True} \\
& \mathrm{Exactly} \left( b_i ^{\left( G \right)}, \, Y_{ij \bm{T}} ^{\left( G \right)}, \;\; \forall j \in \mathcal{I}, \forall \bm{T} \in \mathbb{Z}^3 \right) \hspace{6.75em} \text{($\forall i \in \mathcal{I}$, $\forall G \in \mathcal{G}$)} \\
& \mathrm{AtMost} \left( C_i ^{\left( G \right)}, Y_{ij \bm{T}} ^{\left( G \right)}, \;\; \forall j \in \mathcal{I}, \forall \bm{T} \in \mathbb{Z}^3 \right) \hspace{6.65em} \text{($\forall i \in \mathcal{I}$, $\forall G \in \mathcal{G}$)} \\
& \underset{G \in \mathcal{G}}{\veebar} Y_{ij \bm{T}}^{\left( G \right)} \hspace{19.85em} \text{($\forall i,j \in \mathcal{I}$, $\forall \bm{T} \in \mathbb{Z}^3$)} \\
& \underset{G \in \mathcal{G}}{\vee} \begin{bmatrix}
Y_{ij \bm{T}}^{\left( G \right)} \\
\left| \bm{x}_j + A \bm{T} - \bm{x}_i \right| \le \left( 1 + \varepsilon \right) \left( R_i ^{\left(G \right)} + R_j ^{\left(G \right)} \right)
\end{bmatrix} \hspace{3em} \text{($\forall i,j \in \mathcal{I}$, $\forall \bm{T} \in \mathbb{Z}^3$)} \\
& \neg \Lambda _{ij} ^{\left( G \right)} \implies \neg Y_{ij \bm{T}}^{\left( G \right)} \hspace{15.95em} \text{($\forall i,j \in \mathcal{I}$,  $\forall \bm{T} \in \mathbb{Z}^3$, $\forall G \in \mathcal{G}$)} \\
& \Xi _{ij} ^{\left( G \right)} \, \implies \, Y_{ij \bm{T}}^{\left( G \right)} \hspace{16.95em} \text{($\forall i,j \in \mathcal{I}$,  $\forall \bm{T} \in \mathbb{Z}^3$, $\forall G \in \mathcal{G}$)} \\
& b_i ^{\left( G \right)} \in \mathbb{N}, \quad\; Y_{ij \bm{T}}^{\left( G \right)} \in \left\{ \, \mathrm{True}, \mathrm{False} \, \right\}
\label{eq:boolean-gdp-subproblem}
\end{split}
\end{equation}
\end{widetext}
The set $\mathcal{B}$ is the subset of $G$ corresponding to chemical bonds,
and $0 \le \varepsilon$ is an error.
It is important to note that this subproblem has a similarity with the master problem in the Logic Based Outer-Approximation,
because each atom should form as many chemical bonds with neighboring atoms as possible to minimize the volume of the unit cell when we optimize the Boolean variables $Y_{ij \bm{T}}^{\left( G \right)}$ depending on the spatial arrangement of atoms.
Then we arrive at the objective function of the subproblem being the number of chemical bonds.
It is also important to note that we remove both the lower bounds of feasible coordination numbers and interatomic distances,
because adjacent atoms in a disordered structure can be either very close or very far apart.
This relaxation is reasonable since this subproblem is aimed at creating an opportunity to make a different network of chemical bonds so that all the interatomic distances can be feasible.
Finally,
it is better to memorize the feasibility of the subsets of Boolean variables $Y_{ij \bm{T}}^{\left( G \right)}$ as logic constraints:
\begin{equation}
\mathrm{ML} \left( \bm{Y} \right) = \mathrm{True},
\label{eq:machine-learning-constraint}
\end{equation}
where $\bm{Y}$ is all the set of Boolean variables $Y_{ij \bm{T}}^{\left( G \right)}$.

It is inevitable to solve the two subproblems iteratively many times when an initial structure is randomly generated.
Since the feasible conditions of real variables changes drastically depending on the Boolean variables and vice versa,
well-controlled annealing causes a domino effect to largely transform an initial structure into an optimal solution.
Such an annealing can be realized by applying the steepest-descent method to the subproblem given in Eq.~\eqref{eq:algebraic-gdp-subproblem}:
First, we introduce the linear penalty functions for the constraints on interatomic distances
\begin{equation*}
\begin{split}
d_{\sigma}^{\left(G \right)} \le x_{\sigma} \; &\Rightarrow \; U_{\min} ^{\left(G \right)} \left(x_{\sigma} \right) \equiv \max \left[ 0, k_{\downarrow} \left( d_{\sigma} ^{\left(G \right)} - x_{\sigma} \right) \right] \\
x_{\sigma} \le D_{\sigma}^{\left(G \right)} \; &\Rightarrow \; U_{\max} ^{\left(G \right)} \left(x_{\sigma} \right) \equiv \max \left[ 0, k_{\uparrow} \left( x_{\sigma} - D_{\sigma} ^{\left(G \right)} \right) \right]
\end{split},
\end{equation*}
with $k_{\downarrow}$ and $k_{\uparrow}$ being the constants, and $\sigma$ being the abbreviated index of $ij \bm{T}$.
Accordingly, we have the relaxed optimization problem with $0 < P$ for real variables:
\begin{equation*}
\text{minimize} \quad \sum_{\sigma, G}\left[ U_{\min} ^{\left(G \right)} \left(x_{\sigma} \right) + U_{\max} ^{\left(G \right)} \left(x_{\sigma} \right) \right] + P \left| \det A \right|.
\end{equation*}
The displacements of real variables are calculated from the gradients of the relaxed objective function.
This method is shown to be effective for not only optimizing packings of spheres accurately with small computation~\cite{PhysRevE.103.023307, PhysRevE.104.024101, koshoji2021diverse, PhysRevMaterials.8.113801, 6yl6-fr8b} but also annealing structures without destroying feasible atomic environments too much by adjusting the maximum size of the displacements~\cite{PhysRevMaterials.8.113801, 6yl6-fr8b}.
This annealing is aimed at creating an opportunity to find different spatial arrangements of atoms so that all the Boolean variables can be feasible.
If it succeeds,
the real variables are optimized by gradually decreasing the maximum size of the displacement to verify the feasibility of all the interatomic distances.

The set of Boolean variables $\bm{Y}$ represents the graphs describing crystal structures with clear chemical meanings,
for example,
the subgraphs corresponding to ionic bonds can represent not only coordination polyhedra but also linkings of them,
and similarly,
the central cations of linked polyhedra are joined by the edge corresponding to electrostatic repulsions.
Recall that each edge enforces the associated constraint on the interatomic distance.
As detailed in the previous study~\cite{6yl6-fr8b},
two polyhedra can generally be linked by sharing a common vertex,
a common edge,
or a common face,
but not all of them can be feasible if the central atoms must be far apart.
Similarly,
even if the coordination number is feasible,
the composition of the coordination polyhedron designated by the subgraph is not necessarily feasible.

A considerable amount of subgraphs,
which correspond to coordination polyhedra or linking of them,
are discovered in numerous graphs generated by the solution method.
Since optimizations of real variables with an infeasible graph are very futile,
on-the-fly memorizing of the feasibility of such subgraphs accelerates the discovery of optimal solutions by skipping the precise optimization of real variables with infeasible graphs.
Note that the feasibility of a subgraph can be verified by optimizing the corresponding substructure as discussed in Supplemental Material~\cite{SupplementalMaterial}.
This is why the subproblem given in Eq.~\eqref{eq:boolean-gdp-subproblem} includes the on-the-fly logic constraint given in Eq.~\eqref{eq:machine-learning-constraint},
which consists of the memories of feasible or infeasible subgraphs.

The software \textsf{MARICI} (\textit{Mathematical Architecture for Realizing Inorganic Crystalline materials using mixed-Integer nonlinear programming})~\cite{marici} implements the solution method.
The algorithm is detailed in Supplemental Material~\cite{SupplementalMaterial}.
Numerical performance indicates that the memories of the feasibilities of subgraphs increases the possibility of discovering optimal solutions by at least ten times.
\textsf{MARICI} can efficiently predict crystal structures based on the appropriate assumptions of the atomic radius and feasible coordination numbers of each atom.
For example,
\textsf{MARICI} can design pyroxene and the four superstructures of the perovskite type in about tens of seconds,
and besides,
not only simple ternary oxides such as the spinel but also simple quaternary oxides such as \ch{BaDyFe4O7} in less than ten seconds by only using ten cores of Intel$^{\circledR}$ core$^{\mathrm{TM}}$ Ultra 9 185H (2.30 GHz), 32GB.
\textsf{MARICI} finds not only ionic compounds ($\alpha$-pyrochlore,
quadruple perovskite,
\ch{YBa2Cu3O7},
\ch{NaBe4SbO7},
jarosite without hydrogen,
etc) but also Zintl phases (\ch{MgB2},
\ch{ThSi2},
\ch{AuBe5},
etc) whose crystal structures are composed of both the covalent and ionic bonds.
As detailed in Supplemental Material~\cite{SupplementalMaterial},
the number of optimal solutions of the auxilliary GDP is much smaller than that of the original optimization problem,
which means that the GDP model successfully extracts the meaningful optimal solutions of the original optimization problem.
Finally,
an exhaustive search for quaternary oxides indicates that the GDP model can design many unknown prototypes of crystal structures whose existences in nature would not be surprising:
The prototypes are similar to experimentally-known crystal structures,
especially when they have high symmetries,
but in some cases they belong to minor space groups~\cite{SupplementalMaterial}.

This letter clarifies that the typical form of GDP can naturally formulate the empirical rules of inorganic structural chemistry so that the crystal structures of not only ionic compounds but also Zintl phases can be discovered as local optima.
The efficiency of the solution method indicates that introducing integer variables to extract the ``combinatorial backbone'' of an original continuous optimization problem accelerates visiting every local optimum by iteratively optimizing the continuous and integer variables like the Logic Based Outer-Approximation and remembering infeasible subsets of integer variables to avoid precise optimization of continous variables with infeasible integer variables like the GDP Branch and Bound method.
While many AI techniques have achieved great success~\cite{Merchant2023, Zeni2025},
symmetry-oriented crystal structure prediction~\cite{Zhao2023, FREDERICKS2021107810, Zhang_2025} with the GDP model will discover the majority of known and unknown prototypes of crystal structures based on only the atomic radii and feasible coordination numbers of each atom.
This preprocessing before \textit{ab-initio} simulations broadens the applicability of computational discovery of compounds owing to the efficiency of mathematical crystal chemistry.

\begin{acknowledgements}

This work used computational resources of supercomputer Fugaku provided by the RIKEN Center for Computational Science (Project ID: hp250425, hp260027).
This work also used Fujitsu Server PRIMERGY CX2550 M7 (Miyabi-C) at the Information Technology Center,
The University of Tokyo,
and the facilities (supercomputer Ohtaka and Kugui) of the Supercomputer Center,
The Institute for Solid State Physics,
The University of Tokyo.
This work is supported by JST ASPIRE (Grant No.~JPMJAP2314),
JSPS KAKENHI (Grant No.~26H02231),
The Kazuchika Okura Memorial Foundation,
and Nippon Sheet Glass Foundation for Materials Science and Engineering.
The author is indebted to many researchers including Prof.~Richard Dronskowski,
Prof.~Yoshihiro Kanno,
Prof.~Taisuke Ozaki,
Prof.~Jun-ichi Yamaura,
and Prof.~Yoshihiko Okamoto.
The author has to thank in particular Dr.~Jun Takahashi who not only discussed on NP-hardness but also gave me suggestions of the draft.

\end{acknowledgements}

\clearpage

\setcounter{equation}{0}
\setcounter{table}{0}
\setcounter{figure}{0}

\renewcommand{\theequation}{S\arabic{equation}}
\renewcommand{\thetable}{S-\Roman{table}}
\renewcommand{\thefigure}{S\arabic{figure}}

\onecolumngrid

\begin{center}

\Large \textbf{Supplemental Material for mathematical crystal chemistry}

\vspace{0.05\columnwidth}

\end{center}

\twocolumngrid

\section{Reformulation as Mixed-integer nonlinear programming}

As described in the main article,
a generalized disjunctive programming (GDP) can be reformulated as a mixed-integer nonlinear programming (MINLP).
To do this,
binary variables $y_{ij \bm{T}}^{\left( G \right)} \in \left\{ 0,1 \right\}$, $\xi_{ij \bm{T}}^{\left( G \right)} \in \left\{ 0,1 \right\}$, and $\lambda_{ij \bm{T}}^{\left( G \right)} \in \left\{ 0,1 \right\}$ are introduced to have one-to-one correspondences with the Boolean variables:
$Y_{ij \bm{T}}^{\left( G \right)} = \mathrm{True}$ is equivalent to $y_{ij \bm{T}}^{\left( G \right)} = 1$,
$Y_{ij \bm{T}}^{\left( G \right)} = \mathrm{False}$ is equivalent to $y_{ij \bm{T}}^{\left( G \right)} = 0$,
$\Xi_{ij \bm{T}}^{\left( G \right)} = \mathrm{True}$ is equivalent to $\xi_{ij \bm{T}}^{\left( G \right)} = 1$,
$\Xi_{ij \bm{T}}^{\left( G \right)} = \mathrm{False}$ is equivalent to $\xi_{ij \bm{T}}^{\left( G \right)} = 0$,
$\Lambda_{ij \bm{T}}^{\left( G \right)} = \mathrm{True}$ is equivalent to $\lambda_{ij \bm{T}}^{\left( G \right)} = 1$,
and $\Lambda_{ij \bm{T}}^{\left( G \right)} = \mathrm{False}$ is equivalent to $\lambda_{ij \bm{T}}^{\left( G \right)} = 0$.
It is straightforward to convert logic propositions into algebraic forms using binary variables,
for example,
the logic constraint:
\begin{equation}
\underset{G \in \mathcal{G}}{\veebar} Y_{ij \bm{T}}^{\left( G \right)},
\end{equation}
can be converted into
\begin{equation}
\sum_{G \in \mathcal{G}} y_{ij \bm{T}}^{\left( G \right)} = 1.
\end{equation}
Similarly,
while there are several ways such as Big-M or Hull Reformulation to reformulate the disjunctions:
\begin{equation}
\underset{G \in \mathcal{G}}{\vee} \begin{bmatrix}
Y_{ij \bm{T}}^{\left( G \right)} \\
d_{ij \bm{T}}^{\left(G \right)} \le \left| \bm{x}_j + A \bm{T} - \bm{x}_i \right| \le D_{ij \bm{T}}^{\left(G \right)}
\end{bmatrix},
\end{equation}
the big-M approach can convert the disjunctions into
\begin{align}
& y_{ij \bm{T}}^{\left( G \right)} \, d_{ij \bm{T}}^{\left(G \right)} \le \left| \bm{x}_j + A \bm{T} - \bm{x}_i \right|, \\
&\left| \bm{x}_j + A \bm{T} - \bm{x}_i \right| \le D_{ij \bm{T}}^{\left(G \right)} + M \left( 1 - y_{ij \bm{T}}^{\left( G \right)} \right)
\end{align}
where $M$ is a huge constant.
One of the reformulation of the auxiliary GDP given in Eq.~(4) in the main article is as follows:
\begin{widetext}
\begin{equation}
\begin{split}
\underset{ A, \; \bm{x}_i, \: y_{ij \bm{T}}^{\left( G \right)}}{\text{minimize}} \qquad & \left| \det{A} \right| \\
\text{subject to} \qquad & c_i ^{\left( G \right)} \le \sum_{j \bm{T}} y_{ij \bm{T}} ^{\left( G \right)} \le C_i ^{\left( G \right)} \hspace{15.8em} \text{($\forall i \in \mathcal{I}$, $\forall G \in \mathcal{G}$)} \\
& \sum_{G \in \mathcal{G}} y_{ij \bm{T}}^{\left( G \right)} = 1 \hspace{20em} \text{($\forall i,j \in \mathcal{I}$, $\forall \bm{T} \in \mathbb{Z}^3$)} \\
& y_{ij \bm{T}}^{\left( G \right)} \, d_{ij \bm{T}}^{\left(G \right)} \le \left| \bm{x}_j + A \bm{T} - \bm{x}_i \right| \le D_{ij \bm{T}}^{\left(G \right)} + M \left( 1 - y_{ij \bm{T}}^{\left( G \right)} \right) \hspace{3em} \text{($\forall i,j \in \mathcal{I}$, $\forall \bm{T} \in \mathbb{Z}^3$)} \\
& \xi _{ij} ^{\left( G \right)} \le y_{ij \bm{T}}^{\left( G \right)} \le \lambda _{ij} ^{\left( G \right)} \hspace{17.5em} \text{($\forall i,j \in \mathcal{I}$,  $\forall \bm{T} \in \mathbb{Z}^3$, $\forall G \in \mathcal{G}$)} \\
& d_{ij \bm{T}} ^{\left(G \right)} \equiv r_i ^{\left(G \right)} + r_j ^{\left(G \right)}, \quad\; D_{ij \bm{T}} ^{\left(G \right)} \equiv R_i ^{\left(G \right)} + R_j ^{\left(G \right)} , \quad\; \bm{0} \le A ^{-1} \bm{x}_i \le \bm{1} \\
& A \in \mathbb{R}^3 \times \mathbb{R}^3, \quad\; \bm{x}_i \in \mathbb{R}^3, \quad\; y_{ij \bm{T}}^{\left( G \right)} \in \left\{ \, 0,1 \, \right\}
\label{eq:minlp-crystal-structure-prediction}
\end{split}
\end{equation}
\end{widetext}

\section{Algorithm for solution method}

\begin{figure*}
\centering
\includegraphics[width=2\columnwidth]{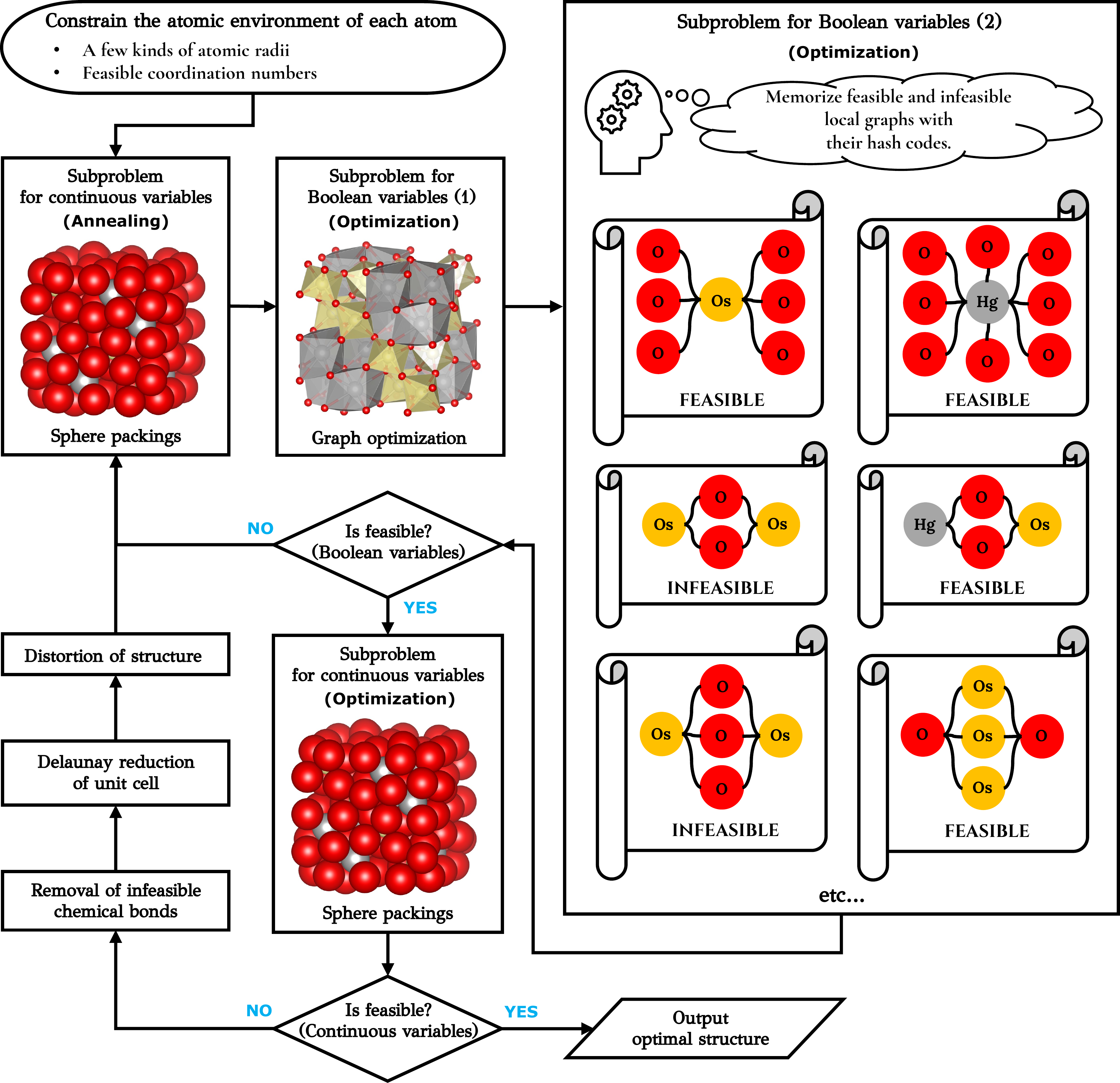}
\caption{
The flowchart of the algorithm for the solution method.
Crystal structures are drawn by \textsf{VESTA} [K. Momma and F. Izumi, Journal of Applied Crystallography \textbf{44}, 1272 (2011)].
}
\label{fig:algorithm}
\end{figure*}

While the main article details the solution method for the auxiliary GDP,
which is equivalent to Eq.~\eqref{eq:minlp-crystal-structure-prediction},
this section details the algorithm for the solution method.
The algorithm is implemented in the software \textsf{MARICI} (\textit{Mathematical Architecture for Realizing Inorganic Crystalline materials using mixed-Integer nonlinear programming}).
The software is available at \url{https://marici.mathematical-crystal-chemistry.solutions},
and it is distributed under the MIT License with its source code for transparent and honest interactions with you.

The algorithm discussed in this section is almost identical to that discussed in the previous studies.
However,
since the source code for \textsf{MARICI} is developed from the scratch to resolve the technical problems,
the computational efficiency is improved.
Besides,
it is found that the on-the-fly constraint comprising the memories of feasibilities of subgraphs also improves the efficiency to find the optimal solutions.
The flowchart of the algorithm is shown in Fig.~\ref{fig:algorithm}.
The two subproblems are iteratively solved to optimize the continuous and Boolean variables with fixing the other.

\subsection{Input data}
\label{sec:input data}

As discussed in the main article,
the atomic environment of each atom is defined by only the atomic radii and the feasible coordination numbers corresponding to the types of interatomic forces:
\begin{equation}
\left( r^{\left( G \right)}, R^{\left( G \right)}, c^{\left( G \right)}, C^{\left( G \right)} \right) \qquad \text{($\forall G \in \mathcal{G}$)},
\end{equation}
where $r^{\left( G \right)}, R^{\left( G \right)} \in \mathbb{R}$ are the minimum and maximum radii,
respectively,
$c^{\left( G \right)}, C^{\left( G \right)} \in \mathbb{N}$ are the minimum and maximum coordination numbers,
respectively,
and $G \in \mathcal{G}$ correponds to the type of interatomic forces.

When we select the chemical compositions for crystal structure prediction,
we must assign the formal charge to each atom (we call a chemical element with formal charge as ``ionic chemical element'').
This is because the formal charges are used for judging whether or not an ionic bond is creatable between atoms as
\begin{equation}
\Lambda _{ij} ^{\left( I \right)} =
\begin{cases}
\mathrm{True} & \text{$q_i q_j < 0$} \\
\mathrm{False} & \text{otherwise}
\end{cases}, \label{eq:ionic_bond_creatability}
\end{equation}
where $I \in G$ corresponds to ionic bonds.
The software \textsf{MARICI} support the five kinds of $G$:
The constraint $I$,
$C$,
$E$,
$I_e$,
$C_e$,
correpond to ionic bonds,
covelent bonds,
electrostatic repulsions,
ionic exclusive repulsions,
and covalent exclusive repulsions,
respectively.
The remained ``creatabilities'' $\Lambda _{ij} ^{\left( G \right)}$ are defined as follows:
\begin{align}
\Lambda _{ij} ^{\left( C \right)} &=
\begin{cases}
\mathrm{False} & \text{$q_i q_j < 0$} \\
\mathrm{True} & \text{otherwise}
\end{cases}, \label{eq:covalent_bond_creatability} \\
\Lambda _{ij} ^{\left( E \right)} &=
\begin{cases}
\mathrm{True} & \text{$q_i q_j > 0$} \\
\mathrm{False} & \text{otherwise}
\end{cases}, \label{eq:electrostatic_repulsion_creatability} \\
\Lambda _{ij} ^{\left( I_e \right)} &=
\begin{cases}
\mathrm{True} & \text{$q_i q_j < 0$} \\
\mathrm{False} & \text{otherwise}
\end{cases}, \label{eq:ionic_exclusion_creataibility} \\
\Lambda _{ij} ^{\left( C_e \right)} &= \mathrm{True}.
\label{eq:covalent_exclusion_creataibility}
\end{align}
\textsf{MARICI} set the exclusive radii as follows:
\begin{equation}
\begin{split}
&r^{\left( I_e \right)} = \chi R^{\left( I \right)}, \\
&r^{\left( C_e \right)} = \chi R^{\left( C \right)}, \\
&R^{\left( I_e \right)} = R^{\left( C_e \right)} = \infty,
\end{split}
\label{eq:exclusive-radii}
\end{equation}
with $\chi = 1.3$.
Accordingly,
when you predict crystal structures with \textsf{MARICI},
the necessary parameters to set feasible atomic environments of each ionic chemical element are given by
\begin{equation}
\left( r^{\left( C \right)}, R^{\left( C \right)}, c^{\left( C \right)}, C^{\left( C \right)}, r^{\left( I \right)}, R^{\left( I \right)}, c^{\left( I \right)}, C^{\left( I \right)}, R^{\left( E \right)} \right).
\end{equation}
If you search only ionic crystal structures,
the necessary parameters are
\begin{equation}
\left( r^{\left( I \right)}, R^{\left( I \right)}, c^{\left( I \right)}, C^{\left( I \right)}, R^{\left( E \right)} \right).
\end{equation}

\subsection{Initial structure generation}
\label{sec:initial_structure_generation}

An initial structure is randomly generated as follows:
Let $\left( a,b,c, \alpha, \beta, \gamma \right)$ be the lattice parameters, and they are initialized randomly under the condition of
\begin{align}
1 \le b,c \le 3, &&
\frac{1}{3} \pi \le \alpha, \beta, \gamma \le \frac{2}{3} \pi,
\end{align}
where $a$ is set to be $1$.
Next, lattice vectors $\left( \bm{a}_1, \bm{a}_2, \bm{a}_3 \right)$ is scaled to satisfy the condition of
\begin{equation}
\frac{1}{\left| \bm{a}_1 \cdot \left( \bm{a}_2 \times \bm{a}_3 \right) \right|} \left[\sum_{i=1} ^M \frac{4}{3} \pi \left( R_i ^{\left( \mathrm{I} \right)} \right)^3 \right] = 0.7.
\end{equation}
Finally, all the atoms are randomly distributed in the unit cell.

\subsection{Subproblem for continous variables}
\label{sec:subproblem-for-continous-variables}

As discussed in the main article,
the solution method for the auxilliary GDP,
which is equivalent to Eq.~\eqref{eq:minlp-crystal-structure-prediction},
consists of the two subproblem.
One of them is the optimization problem for continuous variables,
which is given formulated as Eq.~(7) in the main article.
This GDP can be reformulated as follows:
\begin{equation}
\begin{split}
\underset{ A, \; \bm{x}_i}{\text{minimize}} \qquad & \left| \det{A} \right| \\
\text{subject to} \qquad & y_{ij \bm{T}}^{\left( G \right)} \, d_{ij \bm{T}}^{\left(G \right)} \le x_{ij \bm{T}}, \\
& x_{ij \bm{T}} \le D_{ij \bm{T}}^{\left(G \right)} + M \left( 1 - y_{ij \bm{T}}^{\left( G \right)} \right) \\
& d_{ij \bm{T}} ^{\left(G \right)} \equiv r_i ^{\left(G \right)} + r_j ^{\left(G \right)} \\
& D_{ij \bm{T}} ^{\left(G \right)} \equiv R_i ^{\left(G \right)} + R_j ^{\left(G \right)} \\
& x_{ij \bm{T}} \equiv \left| \bm{x}_j + A \bm{T} - \bm{x}_i \right| \\
& \bm{0} \le A ^{-1} \bm{x}_i \le \bm{1} \\
& A \in \mathbb{R}^3 \times \mathbb{R}^3, \quad\; \bm{x}_i \in \mathbb{R}^3
\end{split}
\end{equation}
The solution method for the continuous optimization algorithm is identical to that discussed in the previous study,
but it is summarized to present this Supplemental Materials in a self-contained manner.
The inequality constraints are approximated by the hard-spherical penalty potential as
\begin{equation*}
\begin{split}
d_{\sigma}^{\left(G \right)} \le x_{\sigma} \; &\Rightarrow \; U_{\min} ^{\left(G \right)} \left(x_{\sigma} \right) \equiv \max \left[ 0, k_{\downarrow} \left( d_{\sigma} ^{\left(G \right)} - x_{\sigma} \right) \right] \\
x_{\sigma} \le D_{\sigma}^{\left(G \right)} \; &\Rightarrow \; U_{\max} ^{\left(G \right)} \left(x_{\sigma} \right) \equiv \max \left[ 0, k_{\uparrow} \left( x_{\sigma} - D_{\sigma} ^{\left(G \right)} \right) \right]
\end{split},
\end{equation*}
where $k_{\downarrow}$ and $k_{\uparrow}$ are a common constant for repulsive and attractive forces,
respectively,
and $\sigma$ is the abbreviated index of $ij \bm{T}$.
Accordingly, the objective function of the subbroblem is approximated as
\begin{equation}
\text{minimize} \quad H \left( \bm{x}, A \right),
\end{equation}
where the $H \left( \bm{x}, A \right)$ is defined as
\begin{equation*}
H \left( \bm{x}, A \right) \equiv \sum_{\sigma, G}\left[ U_{\min} ^{\left(G \right)} \left(x_{\sigma} \right) + U_{\max} ^{\left(G \right)} \left(x_{\sigma} \right) \right] + P \left| \det A \right|
\end{equation*}
with $P$ being the pressure.
The continuous variables are optimized by the steepest-descent method as follows:
Let $\Delta \bm{x}_i$ and $\Delta \bm{a}_i$ be the displacements of $\bm{x}_i$ and $\bm{a}_i$.
They are calculated by
\begin{align}
\Delta \bm{x}_i = - \xi_i \frac{\partial H \left( \bm{x}, A \right)}{\partial \bm{x}_i}, && \Delta \bm{a}_i = - \zeta_i \frac{\partial H \left( \bm{x}, A \right)}{\partial \bm{a}_i},
\end{align}
where the constants $\xi_i$ and $\zeta_i$ are scaled to satisfy the condition:
\begin{align}
\left| \Delta \bm{x}_i \right| \le \Delta x_s ^{\left( \mathrm{max} \right)}, && \left| \Delta \bm{a}_i \right| \le \gamma \Delta x_s ^{\left( \mathrm{max} \right)},
\end{align}
with $\Delta x_s ^{\left( \mathrm{max} \right)}$ being the maximum displacement of atoms in $s$-th optimization step and $\gamma$ being a constant.
$\Delta x_s ^{\left( \mathrm{max} \right)}$ is calculated as
\begin{equation}
\Delta x_s ^{\left( \mathrm{max} \right)} = \Delta x_0 ^{\left( \mathrm{max} \right)} \left( \frac{\Delta x_{S} ^{\left( \mathrm{max} \right)}}{\Delta x_0 ^{\left( \mathrm{max} \right)}} \right) ^{\frac{s}{S}},
\end{equation}
where $S$ is the maximum steps of the steepest-descent method.
In this study,
$\gamma$ is given by
\begin{equation}
\gamma = 0.02,
\end{equation}
and the force constants are set to be
\begin{align}
P = 1.0, && k_{\uparrow} = 30.0, && k_{\downarrow} = -100.0.
\end{align}
When the continuous variables are annealed by the subproblem,
the parameters are given by
\begin{align}
S = 25, && \Delta x_{0} ^{\left( \mathrm{max} \right)} = \Delta x_{S} ^{\left( \mathrm{max} \right)} = 0.5.
\end{align}
When the continuous variables are optimized by this subproblem,
we make two steps composed of local and precise structural optimization.
In the former,
the parameters are given by
\begin{align}
S = 2000, && \Delta x_{0} ^{\left( \mathrm{max} \right)} = 0.3, && \Delta x_{S} ^{\left( \mathrm{max} \right)} = 0.05,
\end{align}
while in the latter,
they are given by
\begin{align}
S = 4000, && \Delta x_{0} ^{\left( \mathrm{max} \right)} = 0.1, && \Delta x_{S} ^{\left( \mathrm{max} \right)} = 0.005.
\label{eq:precise-structural-optimization-parameters}
\end{align}
Note that in the local optimization,
if a structure cannot be an optimal solution after the first $2000$ steps,
the structure is optimized again after initializing $\Delta x_s ^{\left( \mathrm{max} \right)}$,
because in some cases,
not all the interatomic distances can converge into the feasible value.
It is also important to note that in the annealing process,
the tracers of every pairs of atoms are updated per $100$ steps,
the unit cell is refined per 200 steps by using the software \textsf{SPGLIB} [K. S. Atsushi Togo and I. Tanaka, Science and Technology of Advanced Materials: Methods \textbf{4}, 2384822 (2024)],
and a structure is randomly distorted largely per $2500$ annealing steps.
To reduce the computational cost,
the criteria to track a pair of atoms is given as follows:
\begin{align}
0 < q_i q_j &\implies \tau \left( R_i ^{\left( I \right)} + R_j ^{\left( I \right)} \right), \\
0 = q_i q_j &\implies \tau \left( R_i ^{\left( C \right)} + R_j ^{\left( C \right)} \right), \\
0 > q_i q_j &\implies \tau \left( r_i ^{\left( E \right)} + r_j ^{\left( E \right)} \right),
\end{align}
where $\tau$ is set to be $4.0$.
When \textsf{MARICI} appraises the feasibility of continuous variables,
\textsf{MARICI} allows the small error for the feasibilities of interatomic distances as
\begin{equation}
\left( 1 - \varepsilon \right) d_{ij \bm{T}}^{\left(G \right)} \le \left| \bm{x}_j + A \bm{T} - \bm{x}_i \right| \le \left( 1 + \varepsilon \right) D_{ij \bm{T}}^{\left(G \right)},
\end{equation}
where $0 \le \varepsilon$ is a small value.
Since the structural optimization with parameters given in Eq.~\eqref{eq:precise-structural-optimization-parameters} is not so accurate,
\textsf{MARICI} sets $\varepsilon$ as
\begin{equation}
\varepsilon = 0.05
\end{equation}

\subsection{Subproblem for Boolean variables}

\begin{table*}
\caption{
The definitions of model atoms
}
\label{table:model-atoms}
\begin{ruledtabular}
\begin{tabular}{ccc}
Symbol & Feasible formal charges & Feasible atomic environment $\left( r^{\left( C \right)}, R^{\left( C \right)}, c^{\left( C \right)}, C^{\left( C \right)}, r^{\left( I \right)}, R^{\left( I \right)}, c^{\left( I \right)}, C^{\left( I \right)}, R^{\left( E \right)} \right)$ \\ \hline
\ch{O} & $2-$ & $\left( 0, \; 0, \; 0, \; 0, \; 1.40, \; 1.40, \; 0, \; \infty, \; 1.40 \right)$ \\
\ch{Aa} & $1-$ & $\left( 1.1, \; 1.3, \; 2, \; 2, \; 1.80, \; 1.80, \; 0, \; \infty, \; 1.80 \right)$ \\
\ch{Ab} & $2-$ & $\left( 1.1, \; 1.3, \; 3, \; 3, \; 1.80, \; 1.80, \; 0, \; \infty, \; 1.80 \right)$ \\
\ch{Ea} & $1+, 2+, 3+, 4+, 5+, 6+$ & $\left( 0, \; 0, \; 0, \; 0, \; 0.30, \; 0.35, \; 4, \; 4, \; 1.40 \right)$ \\
\ch{Eb} & $1+, 2+, 3+, 4+, 5+, 6+$ & $\left( 0, \; 0, \; 0, \; 0, \; 0.30, \; 0.35, \; 4, \; 4, \; 1.60 \right)$ \\
\ch{Ec} & $2+$ & $\left( 0, \; 0, \; 0, \; 0, \; 0.60, \; 0.60, \; 4, \; 4, \; 1.40 \right)$ \\
\ch{Ed} & $2+$ & $\left( 0, \; 0, \; 0, \; 0, \; 0.60, \; 0.60, \; 5, \; 5, \; 1.40 \right)$ \\
\ch{Ee} & $2+$ & $\left( 0, \; 0, \; 0, \; 0, \; 0.60, \; 0.60, \; 6, \; 6, \; 0.90 \right)$ \\
\ch{Ef} & $1+, 2+, 3+, 4+, 5+, 6+$ & $\left( 0, \; 0, \; 0, \; 0, \; 0.60, \; 0.60, \; 6, \; 6, \; 1.40 \right)$ \\
\ch{Eg} & $1+, 2+, 3+, 4+, 5+, 6+$ & $\left( 0, \; 0, \; 0, \; 0, \; 0.60, \; 0.60, \; 6, \; 6, \; 1.60 \right)$ \\
\ch{Eh} & $4+$ & $\left( 0, \; 0, \; 0, \; 0, \; 0.60, \; 0.70, \; 6, \; 6, \; 1.40 \right)$ \\
\ch{Ei} & $3+$ & $\left( 0, \; 0, \; 0, \; 0, \; 0.80, \; 1.00, \; 8, \; 8, \; 1.60 \right)$ \\
\ch{Ej} & $1+$ & $\left( 0, \; 0, \; 0, \; 0, \; 0.80, \; 1.30, \; 8, \; 8, \; 1.60 \right)$ \\
\ch{Ek} & $2+$ & $\left( 0, \; 0, \; 0, \; 0, \; 1.20, \; 1.40, \; 10, \; 10, \; 1.60 \right)$ \\
\ch{El} & $1+, 2+, 3+$ & $\left( 0, \; 0, \; 0, \; 0, \; 1.40, \; 1.40, \; 12, \; 12, \; 1.80 \right)$ \\
\ch{Em} & $1+, 2+, 3+$ & $\left( 0, \; 0, \; 0, \; 0, \; 1.40, \; 1.60, \; 12, \; 12, \; 1.80 \right)$ \\
\ch{En} & $2+$ & $\left( 0, \; 0, \; 0, \; 0, \; 1.40, \; 1.60, \; 12, \; 7, \; 1.80 \right)$ \\
\ch{Eo} & $2+$ & $\left( 0, \; 0, \; 0, \; 0, \; 1.40, \; 1.80, \; 12, \; 12, \; 1.80 \right)$
\end{tabular}
\end{ruledtabular}
\end{table*}

\begin{table*}
\caption{
The compositions and the number of initial structure generations to design crystal structures.
The number of discovered optimal solutions and the discovered experimentally known crystal structures are also listed.
}
\label{table:crystal-structure-prediction}
\begin{ruledtabular}
\begin{tabular}{cccc}
Composition & Number of initial structures & Number of discovered solutions & Experimental crystal structures \\ \hline
\ch{Ea2Ef4O8} & $1 \times 10^{3}$ & 13 & spinel \\ \hline
\ch{Ea4Ef8O16} & $1 \times 10^{5}$ & 340 & \begin{tabular}{l}
spinel structure \\
$\beta$-\ch{Mg2SiO4}
\end{tabular} \\ \hline
\ch{Ee4El4O12} & $5 \times 10^{3}$ & 4 & \begin{tabular}{l}
Perovskite \\
\ch{BaNiO3} \\
\ch{BaMnO3} \\
\ch{Cs2NaCrO6}
\end{tabular} \\ \hline
\ch{Eg4Ej4O14} & $3 \times 10^{4}$ & 1 & $\alpha$-pyrochlore \\ \hline
\ch{Eh4Ei4O12} & $1 \times 10^{5}$ & 2 & pyroxene \\ \hline
\ch{Ec3Ef4ElO12} & $1 \times 10^{5}$ & 31 & quadruple perovskite \\ \hline
\ch{EaEf4ElO7} & $1 \times 10^{5}$ & 1 & \ch{YBaFe4O7} \\ \hline
\ch{Ea2Ef8El2O14} & $3 \times 10^{5}$ & 2 & \begin{tabular}{l}
\ch{YBaFe4O7} \\
\ch{NaBe4SbO7}
\end{tabular} \\ \hline
\ch{Eb2Eg3ElO14} & $2 \times 10^{5}$ & 22 & jarosite without hydrogen \\ \hline
\ch{EcEd2EiEk2O7} & $1 \times 10^{5}$ & 2 & \ch{YBa2Cu3O7} \\ \hline
\ch{En6Aa6} & $5 \times 10^{3}$ & 5 & \begin{tabular}{l}
\ch{CaSi} \\
\end{tabular} \\ \hline
\ch{Eo4Ab8} & $2 \times 10^{4}$ & 15 & \begin{tabular}{l}
\ch{MgB2} \\
\ch{ThSi2} \\
\ch{AuBe5}
\end{tabular} \\
\end{tabular}
\end{ruledtabular}
\end{table*}

The other subproblem is the optimization problem for Boolean variables,
which is formulated as Eq.~(8) in the main article.
This GDP can be reformulated as follows:
\begin{widetext}
\begin{align}
\underset{y_{ij \bm{T}}^{\left( G \right)}}{\text{maximize}} \qquad & \sum_{i} \sum_{j \bm{T}} \sum_{G \in \mathcal{B}} y_{ij \bm{T}} ^{\left( G \right)} \\
\text{subject to} \qquad & \mathrm{ML} \left( \bm{y} \right) = \mathrm{True} \label{eq:memory-constraint} \\
& \sum_{j \bm{T}} y_{ij \bm{T}}^{\left( G \right)} \le C_i ^{\left( G \right)} \hspace{20.7em} \text{($\forall i,j \in \mathcal{I}$, $\forall \bm{T} \in \mathbb{Z}^3$)} \label{eq:coordination-number-constraint} \\
& \sum_{G \in \mathcal{G}} y_{ij \bm{T}}^{\left( G \right)} = 1 \hspace{22em} \text{($\forall i,j \in \mathcal{I}$, $\forall \bm{T} \in \mathbb{Z}^3$)}  \label{eq:exclusive-constraint} \\
& \left| \bm{x}_j + A \bm{T} - \bm{x}_i \right| \le \left( 1 + \varepsilon \right) \left(R_i ^{\left(G \right)} + R_j ^{\left(G \right)} \right) + M \left( 1 - y_{ij \bm{T}}^{\left( G \right)} \right) \hspace{2.5em} \text{($\forall i,j \in \mathcal{I}$, $\forall \bm{T} \in \mathbb{Z}^3$)} \label{eq:interaction_threshold} \\
& \xi _{ij} ^{\left( G \right)} \le y_{ij \bm{T}}^{\left( G \right)} \le \lambda _{ij} ^{\left( G \right)} \hspace{19.5em} \text{($\forall i,j \in \mathcal{I}$,  $\forall \bm{T} \in \mathbb{Z}^3$, $\forall G \in \mathcal{G}$)} \label{eq:logic-constraint} \\
& y_{ij \bm{T}}^{\left( G \right)} \in \left\{ \, 0,1 \, \right\}
\end{align}
\end{widetext}
Equation \eqref{eq:memory-constraint} is the logic constraint consisting of the memories of feasible or infeasible subgraphs,
where $\bm{y}$ is all the set of binary variables $y_{ij \bm{T}}^{\left( G \right)}$.
Since this integer programming is essentially the logic-based optimization under the logic constraints,
it is easy to solve as follows:
If a constraint $G$ would be enforced between a pair of atoms,
the interatomic distance must be feasible as written in Eq.~\eqref{eq:interaction_threshold}.
This equation determines the creatable chemical bonds,
since the maximum atomic radius of each atom is finite only if $G$ correponds to one of chemical bonds.
To maximize the total number of chemical bonds,
each atom should create as many creatable chemical bonds with neighboring atoms as possible,
while each atom must follow the constraint on the maximum coordination numbers as given in Eq.~\eqref{eq:coordination-number-constraint} and the logic constraints given in Eqs.~\eqref{eq:exclusive-constraint} and \eqref{eq:logic-constraint}.
If the number of creatable chemical bonds is more than the maximum coordination number,
\textsf{MARICI} selects shorter chemical bonds as many as possible.
Finally,
inappropriate chemical bonds are removed to satisfy the constraint of Eq.~\eqref{eq:memory-constraint}.

As discussed in the main article,
two polyhedra can generally be linked by sharing a common vertex,
a common edge,
or a common face,
but not all of them can be feasible if the central atoms must be far apart.
Similarly,
even if the coordination number is feasible,
the composition of the coordination polyhedron designated by the subgraph is not necessarily feasible.
Since Eq.~\eqref{eq:memory-constraint},
which is not easy to formulate,
remembers not only all the infeasible coordination polyhedra of each atom but also all the infeasible linking of coordination polyhedra,
we can easily find them all from a structure.
An infeasible coordination polyhedra or linking of them can be made feasible if we randomly remove chemical bonds which causes infeasibility.

Techinically,
\textsf{MARICI} memorizes all the feasible and infeasible subgraphs corresponding to coordination polyhedra of each atom or linking of coordination polyhedra.
\textsf{MARICI} uses the hash table to remember them:
The key is a subgraph, and the value is its feasibility.
To define the equality of subgraphs,
atoms are sorted with respect to such as atomic numbers and formal charges when creating a subgraph.
The hash code can be generated from the collections of which atoms are linked with each atom.
Each time \textsf{MARICI} finds a subgraph that is not memorized,
\textsf{MARICI} extracts the substructure corresponding to the subgraph to judge whether it is feasible or not.
Accordingly,
the on-the-fly constraint given in Eq.~\eqref{eq:memory-constraint} is regularly updated through crystal structure prediction.
The structural optimization for a substructure is executed by the same method discussed in Sec.~\ref{sec:subproblem-for-continous-variables}.
The parameters for local optimization are given by
\begin{align}
S = 2000, && \Delta x_{0} ^{\left( \mathrm{max} \right)} = 0.2, && \Delta x_{S} ^{\left( \mathrm{max} \right)} = 0.1,
\end{align}
while those for precise optimization are given by
\begin{align}
S = 10000, && \Delta x_{0} ^{\left( \mathrm{max} \right)} = 0.1, && \Delta x_{S} ^{\left( \mathrm{max} \right)} = 0.001.
\end{align}
The feasible error $\varepsilon$ is set to be $0.01$.
Note that random fluctuation is applied to the substructure before structural optimization,
and the operation is repeated five to ten times until a feasible substructure is discovered.
If it fails,
the substructure is infeasible.

\section{Crystal structure prediction}

\begin{table}
\caption{
The number of randomly generated initial structures depending on the number of atoms per unit cell.
}
\label{table:initial-structure-generation}
\begin{ruledtabular}
\begin{tabular}{cc}
Number of atoms per unit cell & Number of initial structure \\ \hline
$1 \sim 10$ & $1 \times 10^{4}$ \\
$11 \sim 15$ & $5 \times 10^{4}$ \\
$16 \sim 20$ & $1 \times 10^{5}$ \\
$21 \sim 25$ & $2 \times 10^{5}$
\end{tabular}
\end{ruledtabular}
\end{table}

\begin{figure*}
\centering
\includegraphics[width=2\columnwidth]{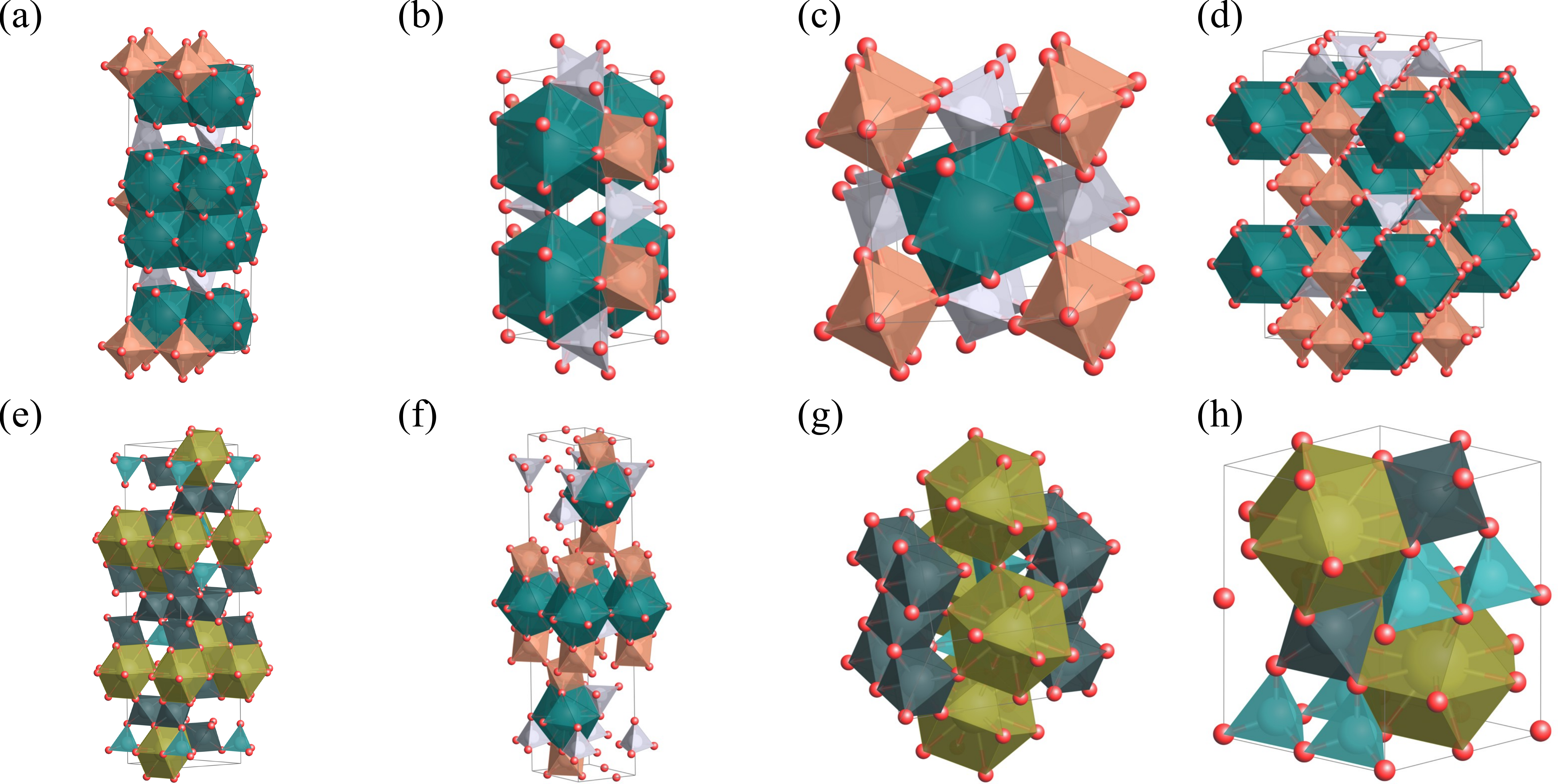}
\caption{
The prototypes of crystal structures designed by \textsf{MARICI}.
Crystal structures are drawn by \textsf{OpenMX Viewer} [Y.-T. Lee and T. Ozaki, Journal of Molecular Graphics and Modelling \textbf{89}, 192 (2019)].
(a) The optimal solution of \ch{EbEgEm2O8}.
It belongs to the minor space group of $Ccc2$ (No.~37).
(b) The optimal solution of \ch{Eb2EgEmO7}.
It belongs to the minor space group of $Pccm$ (No.~49).
(c) The optimal solution of \ch{Eb2EgEmO9}.
It belongs to the minor space group of $P4$ (No.~75).
(d) The optimal solution of \ch{Eb4Eg2EmO18}.
It belongs to the minor space group of $I4$ (No.~79).
(e) The optimal solution of \ch{Ea2Ef5El2O14}.
It belongs to the space group of $R \overline{3}m$ (No.~166).
It corresponds to the crystal structure of \ch{Ba2Cr7O14}.
(f) The optimal solution of \ch{Eb2Eg2EmO11}.
It belongs to the space group of $R32$ (No.~155).
It does not correpond to any experimentally known crystal structures,
but it is similar to the the crystal structure of \ch{BaNb2V2O11}.
(g) The optimal solution of \ch{EaEf2El3O11}.
It belongs to the space group of $P \overline{3}m1$ (No.~164).
It does not correpond to any experimentally known crystal structures,
but it is similar to the the crystal structure of \ch{BaV2Zn3O10H2}.
(h) The optimal solution of \ch{Ea3EfElO7}.
It belongs to the space group of $P6_3 mc$ (No.~186).
It does not correpond to any experimentally known crystal structures,
but it is similar to the the crystal structure of \ch{NaBe4SbO7} which is the optimal solution of \ch{Ea2Ef8El2O14}.
}
\label{fig:crystals}
\end{figure*}

\subsection{Prediction of experimentally known crystal structures}

Table \ref{table:model-atoms} lists the model atoms defined by the feasible formal charges,
the feasible coordination numbers,
and the atomic radii.
Note that both the coordination numbers and atomic radii corresponding to covalent bonds are meaningless when we search ionic crystal structures.
Table \ref{table:crystal-structure-prediction} lists the results of example prediction for the experimentally known crystal structures,
which shows that mathematical crystal chemistry can finds a wide variety of crystal structures.
It is important to emphasis that \textsf{MARICI} can design the crystal structures of Zintl phases consisting of both the covalent and ionic bonds.
Besides,
it is also important to note that the number of local optima is much smaller than that of the original continuous optimization problem.
In some cases,
an experimentally known crystal structure is obtained as the unique optimal solution.
These results indicates that the optimal solutions of the GDP model represents the meaningful optimal solutions of the original optimization problem.
Note that the number of optimal solutions for discovering the quadruple peroskite is much larger than that in the previous study~\cite{6yl6-fr8b} due to larger ionic radius of \ch{Ei}.

There are two optimal solutions for \ch{EcEd2EiEk2O7},
but both of them correspond to the \ch{YBa2Cu3O7} structure;
one of the optimal solutions is a distorted \ch{YBa2Cu3O7} structure.
While \ch{Ec} and \ch{Ed} correspond to the coppers whose coordination numbers are four and five,
respectively,
the coordination polyhedra are very similar;
\ch{Ec} create a square and \ch{Ed} create a square pyramid.
Then,
\ch{YBa2Cu3O7} structure should be discovered as the optimal solution for $\ch{Ec3EiEk2O7}$,
but it is impossible.
This is due to the definition of the ionic exclusive radius $r^{\left( I_e \right)}$;
\textsf{MARICI} uniformly defines it as Eq.~\eqref{eq:exclusive-radii}.
If the radii could be arranged depending on each ionic chemical element,
\ch{YBa2Cu3O7} structure would be discovered as the optimal solution for $\ch{Ec3EiEk2O7}$.

Finally,
the computational efficiency for searching crystal structures is very high,
for example,
not only simple ternary oxides such as the spinel but also simple quaternary oxides such as \ch{YBaFe4O7},
are discovered in less than ten seconds by just using ten cores of Intel$^{\circledR}$ core$^{\mathrm{TM}}$ Ultra 9 185H (2.30 GHz), 32GB.
Besides,
all the optimal solution of \ch{Ee4Ek4O12} corresponding to the experimentally known crystal structures,
which are Perovskite,
\ch{BaNiO3},
\ch{BaMnO3},
and \ch{Cs2NaCrO6} structures,
are discovered in tens of seconds.
Furthermore,
$\alpha$-pyrochlore and pyroxene is discovered in about less than thirty to sixty seconds,
where the flexible ionic radius of \ch{Ej} in the structure of $\alpha$-pyrochlore demands larger computational cost.
On the other hand,
quadruple perovskite and \ch{YBa2Cu3O7} is difficult to find due to the square coordinations of \ch{Ec}.
The ionic radius of \ch{Ec} is large enough to form octahedra as \ch{Ee}~\cite{PhysRevMaterials.8.113801, 6yl6-fr8b},
but copper can create only four ionic bonds.
Besides,
crystal structure prediction of Zintl phases generally demands larger computational costs than that of ionic compounds.

\subsection{Exhaustive search for crystal structures}

This work shortly reports the result of exhaustive search for crystal structures.
More detailed results will be given in future works.
We focus on the two chemical system:
One is $\mathrm{Ea}_k ^{p} \mathrm{Ef}_l ^{q} \mathrm{El}_m ^{r} \mathrm{O}_{n} ^{2-}$ and the other is $\mathrm{Eb}_k ^{p} \mathrm{Eg}_l ^{q} \mathrm{Em}_m ^{r} \mathrm{O}_{n} ^{2-}$,
where the feasible compositions are constrained as:
\begin{equation}
\begin{split}
&1 \le p,q \le 6, \\
& 1 \le r \le 3, \\
& pk + ql + rm - 2n \in \left\{ 0,1 \right\} \\
&1 \le k + l + m + n \le 25, \\
& \frac{1}{7} \le \frac{l}{k}, \frac{m}{k}, \frac{m}{l} \le 7.
\end{split}
\end{equation}
Accordingly,
we obtain the 3055 kinds of compositions per each chemical system.
The number of randomly generated initial structures depending on the number of atoms per unit cell is listed in Table \ref{table:initial-structure-generation}.
If we remove the optimal solutions whose space group numbers are between $1$ to $9$,
the numbers of the discovered optimal solutions of $\mathrm{Ea}_k ^{p} \mathrm{Ef}_l ^{q} \mathrm{El}_m ^{r} \mathrm{O}_{n} ^{2-}$ and $\mathrm{Eb}_k ^{p} \mathrm{Eg}_l ^{q} \mathrm{Em}_m ^{r} \mathrm{O}_{n} ^{2-}$ are $4227$ and $689$,
respectively.

Among the $230$ kinds of space groups,
some of them are not commonly realized by experimantally known crystal structures,
while some of the others such as $Fm \overline{3}m$ (No.~225) are very popular.
On the other hand,
through this exhausive search,
\textsf{MARICI} finds many prototypes of crystal structures belonging the minor groups:
Figures \ref{fig:crystals}(a),
(b),
(c),
and (d),
show the optimal solutions of \ch{EbEgEm2O8},
\ch{Eb2EgEmO7},
\ch{Eb2EgEmO9},
and ch{Eb4Eg2EmO18},
respectively,
which belong to the minor space groups of $Ccc2$ (No.~37),
$Pccm$ (No.~49),
$P4$ (No.~75),
and $I4$ (No.~79),
respectively.
They do not correspond to any experimentally known crystal structures.
Since the symmetries of crystal structures are relevant to quantum properties,
we can expect that mathematical crystal chemistry will predict many crystalline inorganic materials with exotic quantum phenomena by designing unknown crystal structures belonging to minor space groups.

\textsf{MARICI} successfully designs the structures corresponding to the experimentally-known complex crystal structures such as \ch{Ba2Cr7O14},
which is the optimal solution of \ch{Ea2Ef5El2O14} shown in Fig.~\ref{fig:crystals}(e).
Besides,
\textsf{MARICI} also designs many unknown structures whose existences in nature would not be surprising,
for example,
Figs \ref{fig:crystals}(f),
(g),
and (h),
show the optimal solutions of \ch{Eb2Eg2EmO11},
\ch{EaEf2El3O11},
and \ch{Ea3EfElO7},
respectively,
which belong to the space groups of $R32$ (No.~155),
$P \overline{3}m1$ (No.~164),
and $P6_3 mc$ (No.~186),
respectively.
They do not correpond to any experimentally known crystal structures,
but they are similar to the crystal structures of \ch{BaNb2V2O11},
\ch{BaV2Zn3O10H2},
and \ch{NaBe4SbO7},
respectively.

These results show promise for the computational discovery of unknown materials based on mathematical crystal chemistry,
however,
we must assign the appropriate chemical elements to each atom site to appraise the stabilities of materials consisting of the designed crystal structures.
It is important to note that mathematical crystal chemistry is a theory to provide preprocessing before \textit{ab-initio} simulations to accelerate prediction of materials consisting of unknown crystal structures.
If we make success to do so,
the designed materials may have exotic quantum phenomena.
In fact,
\ch{Eb2Eg2EmO11} includes the buckled-honeycomb lattice consisting of ochahedra linked by vertex sharings,
and besides,
\ch{EaEf2El3O11} includes the kagome lattice lattice consisting of ochahedra linked by edge sharings.

\end{document}